\documentclass[usegraphicx,usenatbib]{mn2e}

\begin{document}
\title[X-ray properties of low-mass AGN]{X-ray Spectral and Variability Properties of Low-Mass AGN}
\author[Ludlam et al.]{R. M. Ludlam$^1$,
E. M. Cackett$^1$,
K. G\"{u}ltekin$^2$,
A. C. Fabian$^3$,
L. Gallo$^4$, and
\newauthor
G. Miniutti$^5$
\\$^1$ Department of Physics \& Astronomy, Wayne State University, 666 W. Hancock St., Detroit, MI 48201, USA
\\$^2$ Department of Astronomy, University of Michigan, 500 Church St, Ann Arbor, MI 48109-1042, USA
\\$^3$ Institute of Astronomy, University of Cambridge, Madingley Rd,
Cambridge, CB3 0HA
\\$^4$ Department of Astronomy and Physics, Saint Mary's University, 923 Robie Street, Halifax, NS B3H 3C3, Canada
\\$^5$ Centro de Astrobiologia (CSIC-INTA), Dep. de Astrofisica; ESAC, P.O: Box 78, E-28691 Villanueva de la Canada, Madrid, Spain
}
\date{Received ; in original form }
\maketitle

\begin{abstract}
We study the X-ray properties of a sample of 14 optically-selected low-mass AGN whose masses lie within the range $10^{5}-2\times10^{6} M_\odot$ with {\it XMM-Newton}.  Only six of these low-mass AGN have previously been studied with sufficient quality X-ray data, thus, we more than double the number of low-mass AGN observed by {\it XMM-Newton} with the addition of our sample.  We analyze their X-ray spectral properties and variability and compare the results to their more massive counterparts.  The presence of a soft X-ray excess is detectable in all five objects which were not background dominated at $2-3$ keV.  Combined with previous studies, this gives a total of 8 low-mass AGN with a soft excess.  The low-mass AGN exhibit rapid, short-term variability (hundreds to thousands of seconds) as well as long-term variability (months to years).  There is a well-known anti-correlation between black hole mass and variability amplitude (normalized excess variance).  Comparing our sample of low-mass AGN with this relation we find that all of our sample lie below an extrapolation of the linear relation.  Such a flattening of the relation at low masses (below $\sim 10^6$~M$_\odot$) is expected if the variability in all AGN follows the same shape power spectrum with a break frequency that is dependent on mass.  Finally, we also found two objects that show significant absorption in their X-ray spectrum, indicative of type 2 objects, although they are classified as type 1 AGN based on optical spectra.
\end{abstract}


\section{Introduction}
Supermassive black holes (with $M\sim10^6 - 10^9~M_\odot$) are thought to be at the center of  all massive galaxies.  Black hole mass measurements have unveiled tight correlations between the host galaxy properties and the black hole mass \citep[the $M-L$ and $M-\sigma$ relations, e.g.][and references therein]{gultekin09}, and have led to the study of how black holes and galaxies co-evolve \citep[see][for a review]{kormendy13}.  Within this study of co-evolution, there has been much interest in intermediate-mass black holes with $\sim 10^5 - 10^6~M_\odot$, since these objects are both closer to their original primordial mass, as well as existing in smaller, and sometimes bulgeless host galaxies \citep[see][for a review]{greene12}.

An initial sample of 19 intermediate-mass black hole (IMBH) candidates in AGN was determined by \citet{greene04} based on the first SDSS data release.  Using the fourth SDSS data release  \citet{Greene07} significantly increased the sample size to close to 200 IMBHs. The sample was initially selected from a search of all AGNs that had a broad H$\alpha$ emission line.  The black hole masses of this sample were then estimated using the radius-luminosity relation \citep[e.g.][]{kaspi00,vestergaard02,vestergaard06,bentz06,bentz09,bentz13} to obtain an estimate for the radius of the broad line region, and combined with the width of the emission line to estimate the virial black hole mass \citep[described in][]{Greene07}.  The $\sim$200 objects deemed to be within the intermediate mass range ($<2\times10^6~M_\odot$) were then selected for further study.  A fraction of these (including the objects in our sample) were previously detected by \emph{ROSAT} and noted for their soft X-ray luminosity. While short snapshot X-ray observations of many of these objects have been performed with {\it Chandra} \citep{greene07_xray,desroches09,dong12}, thus far, detailed X-ray spectral fitting has only been performed for a sample consisting of only six IMBHs  \citep{dewangan08, miniutti09}. Those objects were chosen for their low BH mass, AGN-like spectrum, as well as their high Eddington ratio.

Two well-known low-mass AGN are the spiral galaxy NGC 4395 and the dwarf elliptical galaxy POX 52. Both are seen to be highly variable.  A study by \citet{Iwasawa00} noted that the rapid and strong X-ray variability of NGC 4395 was consistent with being a scaled-down Seyfert 1 galaxy. In addition, \citet{Moran05} observed a number of X-ray properties that make this source unique among type I AGN such as the X-ray spectral slope changing from $\Gamma<1.25$ to $\Gamma>1.7$ on the timescale of roughly a year. \citet{Thornton08} looked at the spectral and variability properties from observations of POX 52 from \emph{XMM-Newton} and \emph{Chandra}, finding variability both on short, 500s, timescales and the 9 month timescale between the two observations.

Another study by \citet{dewangan08} performed extensive analysis on both NGC 4395 and POX 52. They characterized their variability, analyzed their X-ray spectrum, and examined the UV emission and optical-to-X-ray spectral index. Short timescale variability was seen in both NGC 4395 and POX 52. NGC 4395 was found to be the most variable of all AGNs that were previously studied. \citet{vaughan05} had previously noted that this object's X-ray emission is the most variable of all AGNs and exceeds 100$\%$ in its fractional variability amplitude.  Furthermore, its normalized excess variance was the highest value in comparison with more massive AGNs.  Both NGC 4395 and POX 52 exhibited broad spectral characteristics of Seyfert 1 galaxies, but NGC 4395 lacked soft excess emission that is a known characteristic to narrow-line Seyfert 1s (NLS1s).  From their examination of the data from \emph{XMM-Newton}'s OM, \cite{dewangan08} were again able to confirm that NGC 4395 and POX 52 resemble NLS1s from their optical-to-X-ray spectral energy distributions.  

 \citet{miniutti09} followed with a study of four IMBHs from the original sample of 19 IMBHs by \citet{greene04}: GH 1, GH 8, GH12, and GH 14 \citep[note that when using the ID from][these are GH 11, GH 126, GH 171 and GH 181]{Greene07}.  They focused their research on the X-ray spectral properties, presence of a soft excess, and variability of these lesser known IMBHs.  The spectra of the IMBHs showed the same shape as the spectra of more massive AGNs.  They found that, when fitted with a broken power law, the X-ray spectra of their sample were consistent with the photon indices seen in Palomar-Green (PG) quasars in both the hard (2-10 keV) and soft ($<$ 2 keV) energies. Moreover, three of the four objects showed the presence of a soft excess in their spectra which is seen in more massive Seyfert 1 galaxies and quasars.  This did not shed any more light on the properties and origins of the soft excess emission.  However, these objects were extremely X-ray variable when compared to more massive AGNs, just like NGC 4395 and POX 52.  Their normalized excess variance was higher than the more massive radio-quiet Seyfert 1 galaxies and began to fill in the lower mass portion of the $\sigma_{\rm NXS}^2$--$M_{\rm BH}$ relation.  The normalized excess variance, $\sigma_{\rm NXS}^2$, is a simple measure of the variability amplitude of an object. In the $\sigma_{\rm NXS}^2$--$M_{\rm BH}$ relation, as BH mass increases the normalized excess variance decreases. In addition to beginning to fill in the low mass region, \citet{miniutti09} determined that their sample of IMBH accretion properties were similar to those of more massive AGNs.  This suggests that they are accreting in the same fashion as more massive AGNs. 

In this work, we analyze fourteen additional IMBHs to more than triple the existing sample. The main objectives are to analyze the X-ray spectral properties and characterize the variability of each object.  One of the motivations of this study is to see if the conclusions of Miniutti et al. still hold with the addition of a larger sample. In the spectral analysis  we will fit the data with various models and look for the presence of a soft excess. We will see how the spectra of this sample compare to more massive AGNs, in addition to looking at the variability properties of the sample.  We detail our sample in Section~\ref{sec:sample}, describe the observations in Section~\ref{sec:obs} and our analysis and results in Section~\ref{sec:res}. Finally, we discuss our findings and their implications in Section~\ref{sec:discuss}.

\section{The Sample}\label{sec:sample}
The objects for this sample were based upon the SDSS objects with black hole masses estimated to be less than $2\times10^{6} M_\odot$ that were published in  \citet{Greene07}.  As part of our {\it XMM-Newton} AO10 program (PI: Cackett) we observed 8 IMBHs.  We also searched the archives for other IMBHs with {\it XMM-Newton} observations longer than 10 ks, leading to a total sample size of 14 objects.  Table~\ref{tab:sample} contains the information on the fourteen SDSS objects and their corresponding Greene \& Ho (GH) name which were taken from the identification number in table 1 of \citet{Greene07}. Those objects that had multiple observations were given a letter that correlates to their event file. We analyze the X-ray spectral properties and characterize the variability of these fourteen objects, as well as, compare them to more massive AGN.  Note that objects from \citet{greene04} are also in \citet{Greene07}, but we only use the ID number from \citet{Greene07} to identify the objects.

\begin{table*}
\caption{Information on the sample of observed objects }
\label{tab:sample}
\begin{minipage}{177mm}
\begin{tabular}{lcccccccc}
\hline
\multicolumn{1}{c}{Object}
&\multicolumn{1}{c}{SDSS Name}
&\multicolumn{1}{c}{Event file}
&\multicolumn{1}{c}{Observation Date}
&\multicolumn{1}{c}{z}
&\multicolumn{1}{c}{$N_{\rm H}$}
&\multicolumn{1}{c}{PN exp}
&\multicolumn{1}{c}{MOS1 exp}
&\multicolumn{1}{c}{MOS2 exp}\\
\hline
GH 18
&SDSS J022849.51--090153.7
&0674810101
&17/01/2012
& 0.0722
& 3.43
&8.7
&11.3
&11.3\\
GH 47
&SDSS J082443.28+295923.5
&0504102001
&03/11/2007
&0.0254
&3.57
&19.0
&23.0
&23.0\\
GH 49
&SDSS J082912.67+500652.3
&0303550901
&26/04/2006
&0.0435
&4.05
&11.4
&16.9
&16.9\\
GH 78
&SDSS J094057.19+032401.2
&0306050201
&30/10/2005
&0.0606
&3.45
&21.9
&26.3
&26.3\\
GH 79a*
&SDSS J094240.92+480017.3
&0201470101
&14/10/2004
&0.197
&1.22
&39.5
&50.1
&50.1\\
GH 79b
&
&0201470301
&13/11/2004
&
&
&14.4
&14.3
&14.3\\
GH 91a
&SDSS J102348.44+040553.7
&0108670101
&05/12/2000
&0.0989
&2.78
&45.7
&53.1
&53.1\\
GH 91b
&
&0605540201
&13/12/2009
&
&
&92.6
&119
&120\\
GH 91c*
&
&0605540301
&08/05/2009
&
&
&48.0
&63.6
&63.6\\
GH 94a
&SDSS J103234.85+650227.9
&0400570401
&06/05/2006
& 0.0056
& 1.25
&19.5
&23.5
&23.5\\
GH 94b
&
&0674810701
&10/10/2011
&
&
&12.8
&15.3
&16.4\\
GH 112
&SDSS J111644.65+402635.5
&0674810401
&03/11/2011
& 0.202
& 1.72 
&14.6
&17.8
&18.3\\
GH 116
&SDSS J112333.56+671109.9
&0503600401
&22/05/2007
&0.055
&1.10
&17.2
&19.4
&19.7\\
GH 138a*
&SDSS J115601.13+564923.3
&0674810801
&29/05/2011
& 0.118
& 1.43
&4.6
&---
&---\\
GH 138b
&
&0674811201
&05/11/2011
&
&
&8.8
&11.5
&11.5\\
GH 142a*
&SDSS J122342.81+581446.1
&0505010101
&19/06/2007
& 0.0143
& 1.18
&7.7
&13.8
&13.9\\
GH 142b*
&
&0674810301
&03/05/2011
&
&
&2.6
&---
&---\\
GH 142c*
&
&0674810901
&20/06/2011
&
&
&2.3
&0.6
&0.7\\
GH 142d*
&
&0674811101
&30/10/2011
&
&
&6.1
&8.8
&8.8\\
GH 181a
&SDSS J143450.62+033842.5
&0305920401
&18/08/2005
& 0.0283
& 2.51
&18.4
&26.9
&27.5\\
GH 181b
&
&0674810501
&16/08/2011
&
&
&10.2
&13.1
&13.1\\
GH 211a
&SDSS J162636.40+350242.0
&0505010501
&17/08/2007
& 0.0341
& 1.44
&10.5
&13.4
&13.4\\
GH 211b
&
&0505011201
&19/08/2007
&
&
&14.0
&17.4
&17.4\\
GH 211c
&
&0674810201
&16/08/2011
&
&
&4.7
&7.7
&7.7\\
GH 211d
&
&0674811001
&17/01/2012
&
&
&10.5
&13.6
&13.4\\
GH 213
&SDSS J163159.59+243740.2
&0674810601
&28/08/2011
& 0.0433
& 3.69
&11.5
&14.4
&14.5\\
\hline
\end{tabular}

\medskip
The galactic column density, $N_{\rm H}$, is given in the units of $10^{20}$ cm$^{-2}$.  Net exposure time is given in ks. Observations GH 138a and GH 142b had both EPIC-MOS detectors closed at the time of their observations. * indicates observations that have significant periods of high background that was unable to be eliminated.
\end{minipage}
\end{table*}

\section{Observations and Data Reduction}\label{sec:obs}
All observations of our sample were taken with \emph{XMM-Newton} operating in prime full-window imaging mode with either medium or thin filters. Table~\ref{tab:sample} also includes each object's redshift, galactic column density ($N_{\rm H}$), as well as the exposure time of each EPIC detector.  The exposure times listed are the net exposure times after the data have been filtered for spectral analysis. The X-ray data were reduced by following the step-by-step Science Analysis System (SAS ver. 13.0.0) threads for the EPIC detectors.  Good time interval files were created to filter out periods of background flaring for GH 94b, GH 112, GH 116, and GH 181a. For the others, background filtering was either not needed or too much data would have been lost if it was performed.  The objects that suffered from high background that was unable to be eliminated are denoted with an asterisk in Table~\ref{tab:sample}. The lightcurves only use the X-ray data from the EPIC-PN and were binned to 200s.  Depending on the brightness of the image, we used a circular source extraction region with radius 20--30 arcsec.

\section{Analysis and Results}\label{sec:res}
\subsection{X-ray Spectral Analysis}
The spectral analysis for each of the objects used data from all three detectors and had 25 counts per energy bin to allow the use of $\chi^2$ minimization. The data were analyzed using Xspec (ver. 12.8.0) in the 0.3 -- 10 keV energy range. 

\begin{figure*}
\centering
\includegraphics[angle=270,width=8cm]{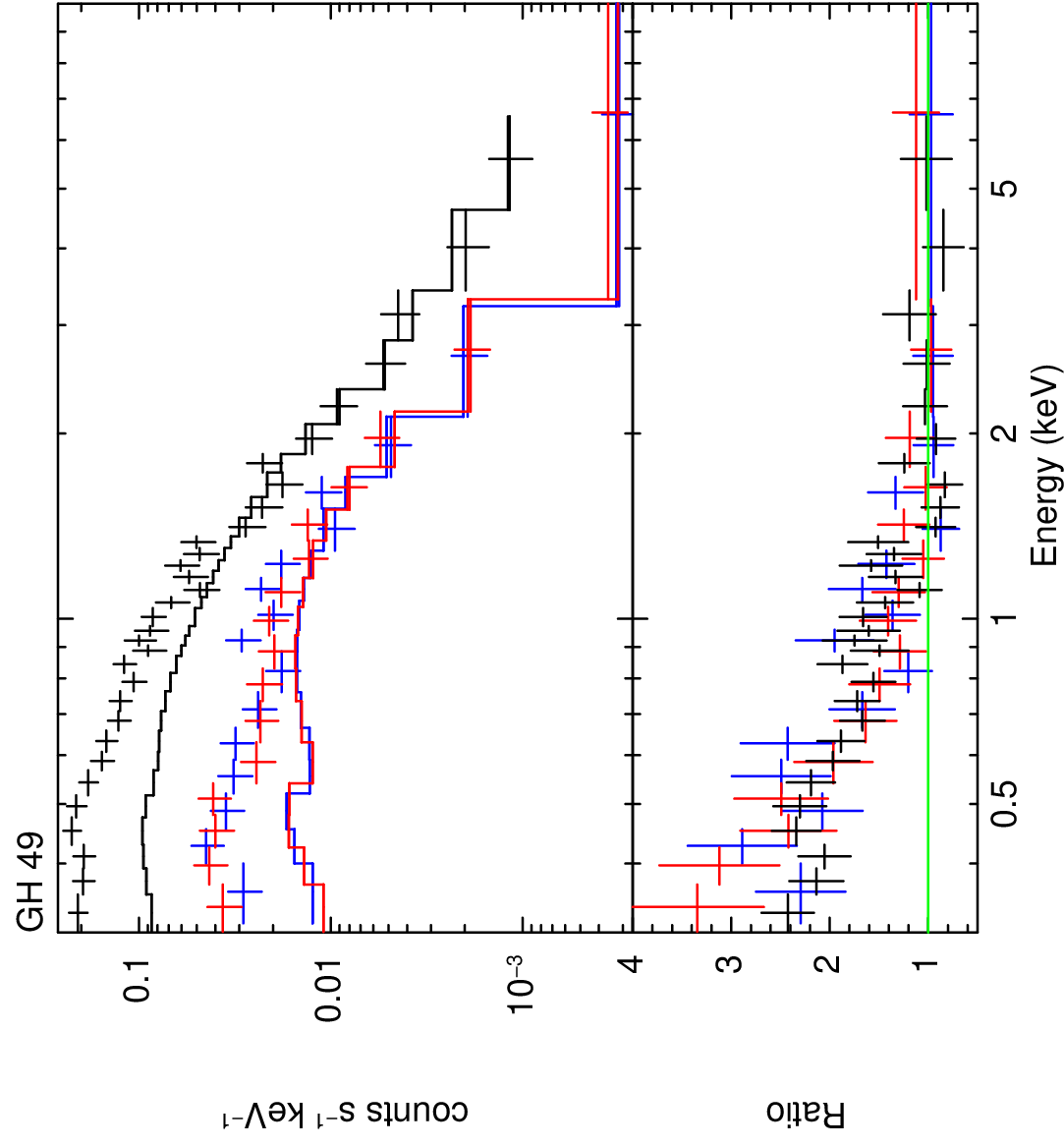}
\hspace{0.5cm}
\includegraphics[angle=270,width=8cm]{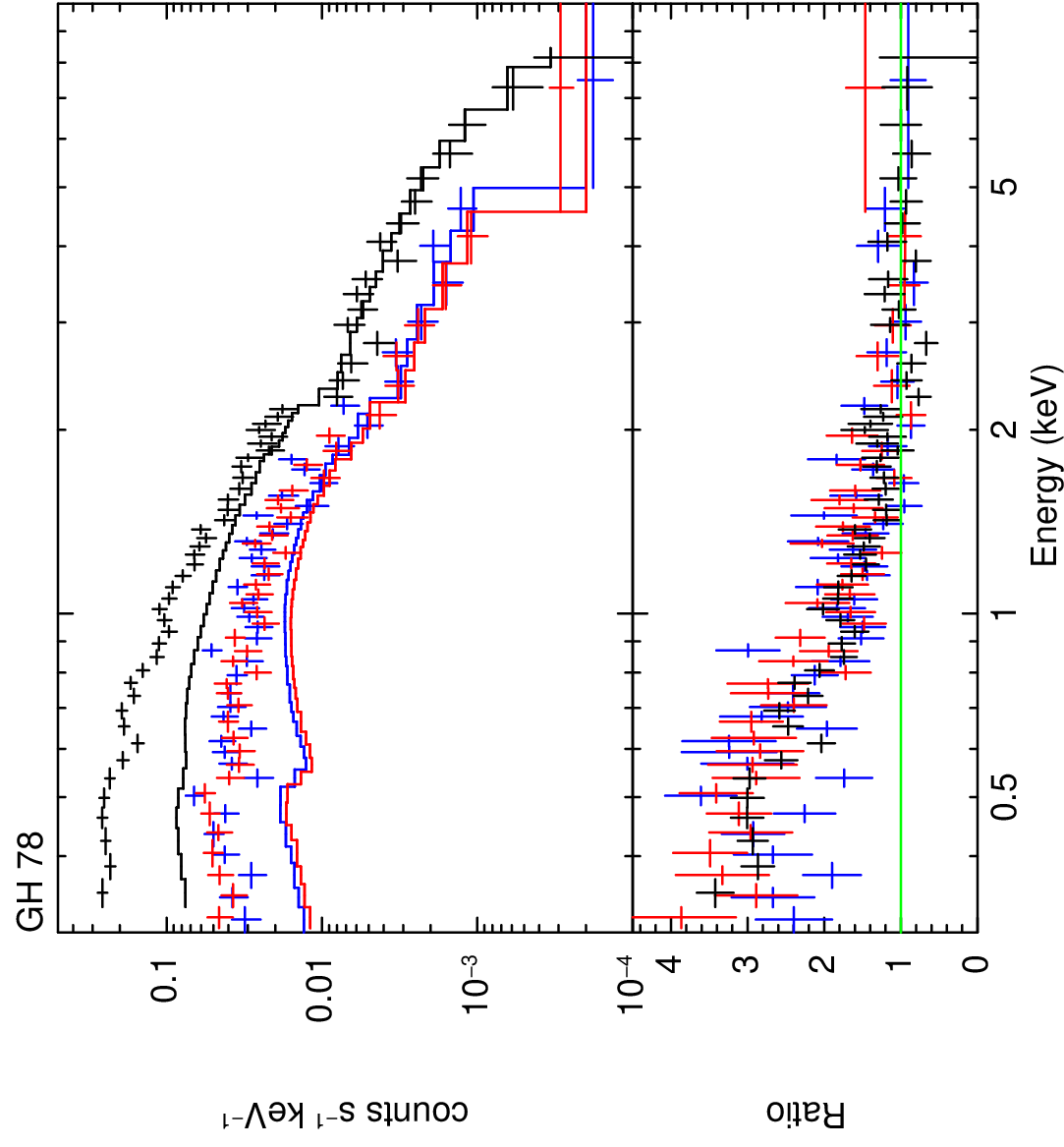}
\includegraphics[angle=270,width=8cm]{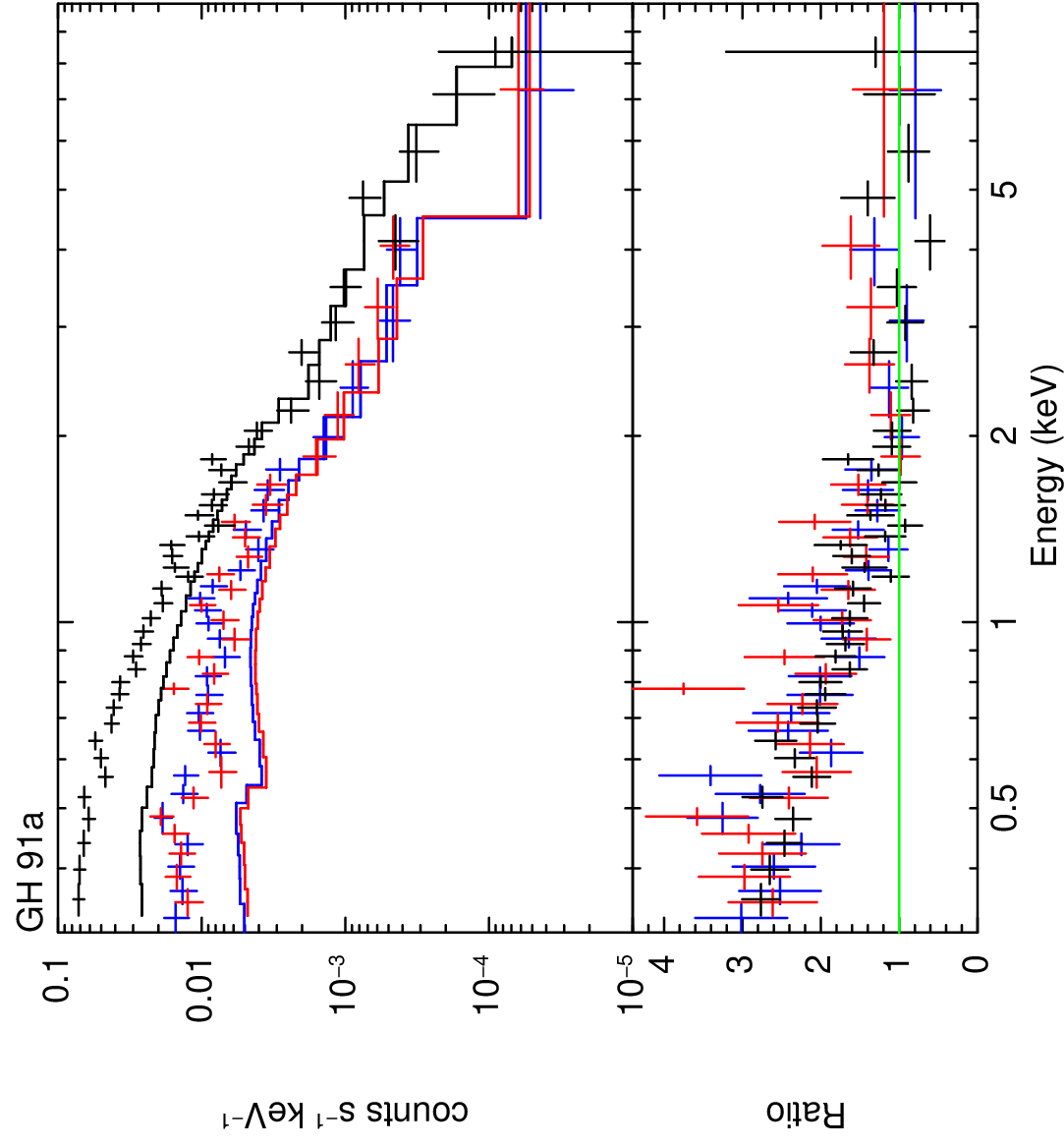}
\hspace{0.5cm}
\includegraphics[angle=270,width=8cm]{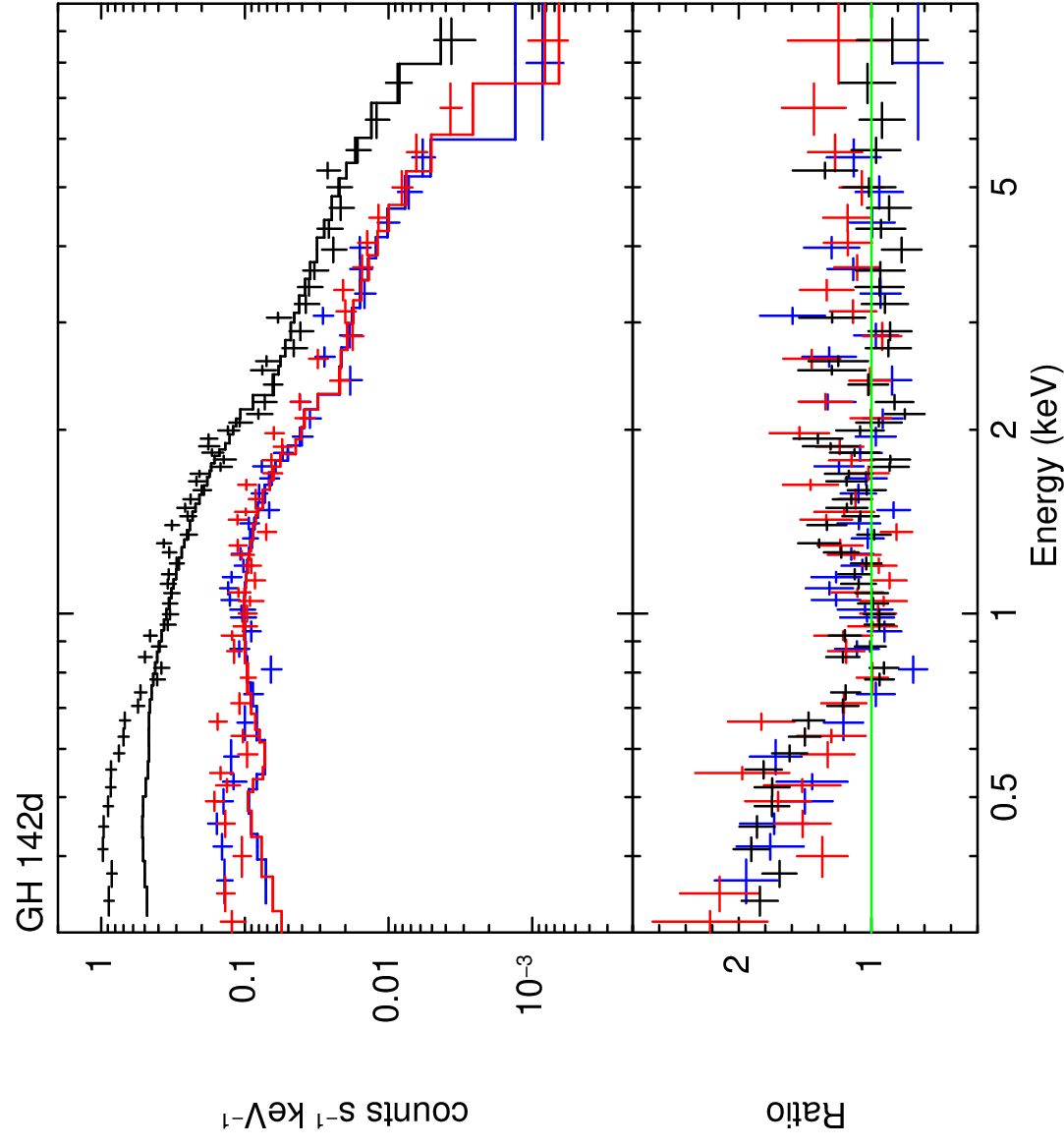}
\includegraphics[angle=270,width=8cm]{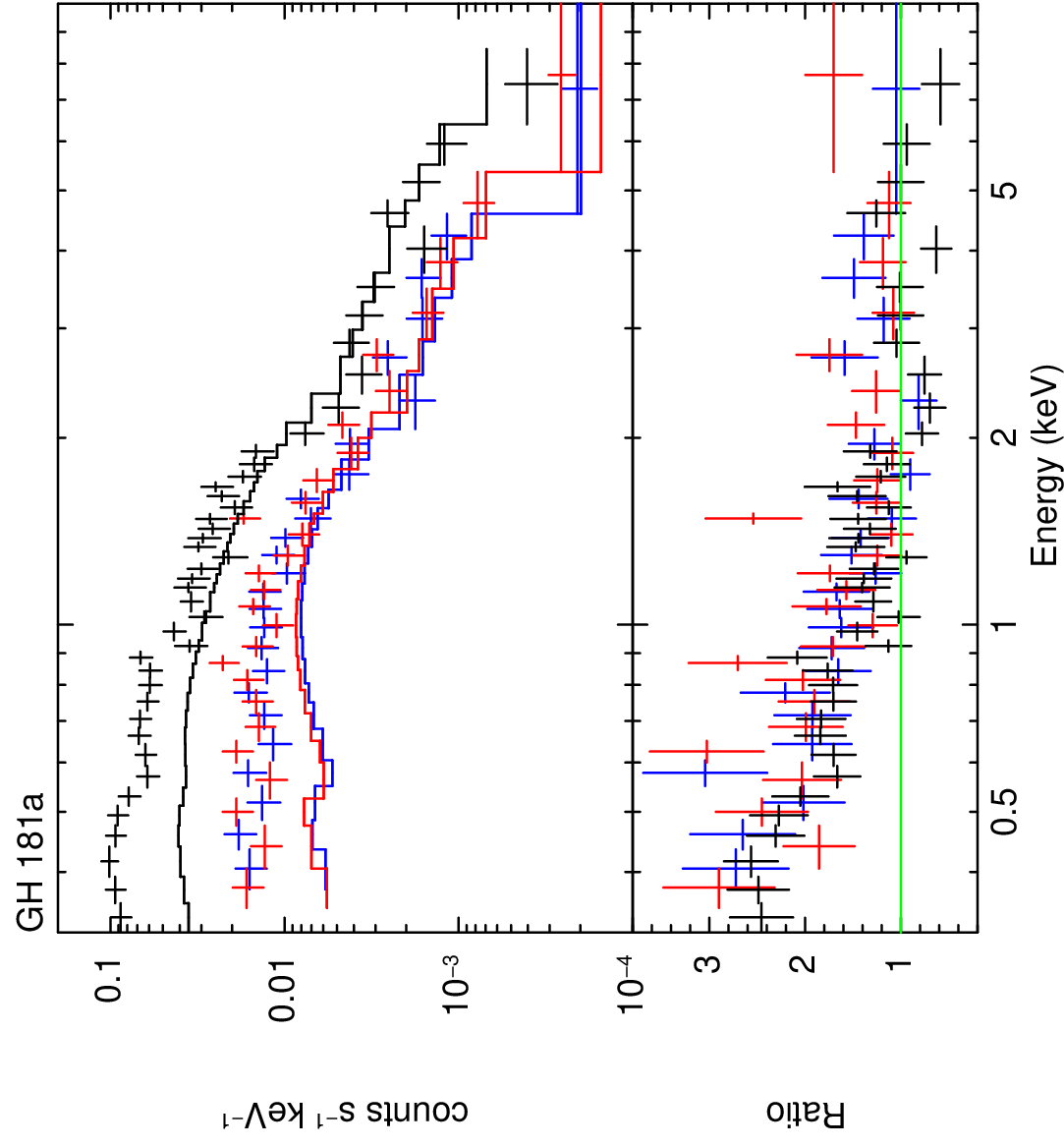}
\caption{Spectra of the five objects that have a soft excess: GH 49, GH 78, GH 91a, GH 142d, and GH 181a. 
The spectra are fit with a power-law above 2 keV only.  The panel below each spectrum shows the ratio of the simple power-law to the data, showing strong deviations below 2 keV, indicating a soft excess.  Blue, red and black points are from the MOS1, MOS2 and PN detectors respectively.}
\label{fig:softx}
\end{figure*}

A characteristic of more massive NLS1s is the presence of a soft excess.  To examine the presence of a soft excess we first fit a simple power-law model to the 2 -- 10 keV spectrum.  We include photoelectric absorption from our own Galaxy with $N_{\rm H}$ fixed at the Galactic column density toward each object.  Many of the PN spectra are background-dominated from 2 -- 3 keV upwards not allowing for a reliable determination of the 2 -- 10 keV spectral slope.  Therefore, for the purposes of looking for the presence of a soft excess, we only examine the 5 sources which have at least one spectrum that is source-dominated at these energies, these are GH 49, GH 78, GH 91, GH 142 and GH 181.  After obtaining a fit to the 2 -- 10 keV spectrum, we extrapolate the fit back to the 0.3 -- 2 keV and look for an excess above the model.  We find that all 5 sources show a soft excess.  We give the 2 -- 10 keV fits in Table~\ref{tab:pow}.  Uncertainties on the parameters are quoted at a 1$\sigma$ confidence level.  The $\chi^2$ value over the 0.3 -- 10 keV region is given to indicate the poor fit when extrapolating the model, and hence the presence of a soft excess.  We also show the soft excesses for these 5 objects in Figure~\ref{fig:softx}.  The addition of these five objects to the IMBHs studied previously by Minuitti et al. give eight objects in this mass range that exhibit a presence of soft excess.  The soft excesses all appear reasonably smooth, and no additional components emission or absorption lines are statistically required.  The photon indices, $\Gamma$, are close to or within the range consistent with those seen in more massive radio quiet AGNs \citep[$1.7<\Gamma<2.6$;][]{dewangan08}.

We also fit all spectra over the 0.3 -- 10 keV band with an absorbed blackbody plus power-law model, where the blackbody is used to characterize the soft excess.  The best fitting parameters are given in Table \ref{tab:bb}.    The blackbody temperatures were typical of NLS1s in the range of $0.1< kT <0.2$ keV \citep{gier04}.  While the high-background above 2 keV in many observations has prevented a direct search for a soft excess in many of the objects, indications of a soft excess can be determined from comparing the power-law index in those objects that we have confirmed have a soft excess with the remaining sources.  The similar power-law index in most sources may indicate that a soft excess could be present in many of the objects, though better data above 2 keV would be required to confirm this.

GH 47 and GH 94 had a different spectral shape than all the other objects, with a spectral shape resembling that of Type II AGN.  GH 47 and GH 94 were omitted from the spectral fit tables since their spectra cannot be described properly by the same models as the Type I objects.  We describe their spectra in more detail in Section~\ref{sec:type2}.

\begin{table*}
\caption{Spectral parameters for power-law fits to the data between $2-10$ keV.}
\label{tab:pow}
\begin{center}
\begin{tabular}{lcccc}
\hline
Object & $\Gamma_{2-10}$ & norm ($10^{-5}$) & $\chi^2_{2-10}$/dof & $\chi^2_{0.3-10}$/dof \\
\hline
GH 18  & $2.11\pm0.26$& $6.04_{-1.66}^{+2.21}$& $0.21$ & $6.17$ \\
GH 78  & $1.82\pm0.11$ & $13.6\pm1.9$ & $0.85$ & $12.15$\\
GH 91a  & $1.96\pm0.20$& $1.90_{-0.42}^{+0.53}$& $1.06$& $8.37$\\
GH 142d  &$1.58\pm0.07$&$35.2_{-2.99}^{+3.24}$&$0.98$&$3.05$ \\
GH 181a  &$1.63\pm0.13$&$3.32\pm0.60$&$1.88$&$5.05$ \\
GH 181b  &$1.79\pm0.28$&$4.17\pm1.77$&$0.87$&$2.94$ \\
\hline
\end{tabular}
\end{center}
\small
Note.--- The Galactic column density, $N_{\rm H}$, is given in Table~\ref{tab:sample} and is fixed during the fits. $\Gamma$ is the power-law index. The normalization of the power-law is defined as the  photon flux (photons keV$^{-1}$ cm$^{-2}$ s$^{-1}$) at 1 keV.
\end{table*}

\begin{table*}
\caption{Spectral parameters for fits with a power-law plus a blackbody.}
\label{tab:bb}
\begin{minipage}{160mm}
\begin{center}
\begin{tabular}{lccccccc}
\hline
Object & $kT$ & norm ($10^{-6}$) & $\Gamma$ & norm ($10^{-5})$ & $\chi^2/dof$ & $F_{0.3-10}$  & $F_{2-10}$  \\
\hline
GH 18 & $0.11 \pm 0.01$ & $1.81 \pm 0.42$ & $2.29 \pm 0.11$ & $7.2 \pm 0.6$ & $39.31/60$ & $3.76 \pm 0.12$ & $1.20 \pm 0.11$  \\
GH 49 & $0.11 \pm 0.01$ & $9.06 \pm 1.01$ & $2.39 \pm 0.05$ & $43.8 \pm 1.4$ & $269.29/276$ & $21.1 \pm 0.3$ & $6.38 \pm 0.25$  \\
GH 78 & $0.12 \pm 0.01$ & $3.37 \pm 0.57$ & $2.15 \pm 0.06$ & $20.2 \pm 1.0$ & $156.55/145$ & $10.8 \pm 0.2$ & $4.17 \pm 0.17$ \\
GH 79a &$0.19 \pm 0.06$ &$0.21 \pm 0.13$ &$2.98 \pm 0.28$ &$2.11 \pm 0.46$&$108.71/96$ &$1.12 \pm 0.05$ &$0.14 \pm 0.06$ \\
GH 79b &$0.11 \pm 0.01$ &$0.92 \pm 0.45$ &$2.05 \pm 0.51$ &$1.76 \pm 0.62$ &$13.05/13$ &$1.39 \pm 0.17$ &$0.42 \pm 0.15$\\
GH 91a &$0.12 \pm 0.01$ & $0.40 \pm 0.12$ & $2.27 \pm 0.09$ & $2.75 \pm 0.19$ & $81.62/97$ & $1.14 \pm 0.03$ & $0.48 \pm 0.03$  \\
GH 91b &$0.13 \pm 0.01$ & $0.59 \pm 0.12$ & $2.39 \pm 0.06$ & $3.98 \pm 0.17$ & $216.2/231$ & $1.98 \pm 0.04$ & $0.58 \pm 0.04$   \\
GH 112 &$0.12 \pm 0.01$ & $2.1 \pm 0.5$ & $2.7 \pm 0.1$ & $7.4 \pm 0.5$ & $89.28/100$ & $4.23 \pm 0.08$ & $0.67 \pm 0.07$  \\
GH 116 & $0.16 \pm 0.07$ & $0.27 \pm 0.20$ & $2.27 \pm 0.35$ & $1.47 \pm 0.42$ & $23.48/17$ & $0.87 \pm 0.10$ & $0.25 \pm 0.09$  \\
GH 138b &$0.15 \pm 0.02$ & $1.04\pm 0.43$ & $2.47 \pm 0.11$ & $7.81 \pm 0.67$ & $90.19/86$ & $4.12 \pm 0.12$ & $1.01 \pm 0.12$  \\
GH 142a &$0.08 \pm 0.01$ & $6.55 \pm 0.68$ & $1.69 \pm 0.03$ & $42.7 \pm 0.8$ & $377.20/328$ & $31.3 \pm 0.6$ & $17.6 \pm 0.5$  \\
GH 142b &$0.07 \pm 0.01$ & $6.8 \pm 2.2$ & $1.70 \pm 0.08$ & $46.3 \pm 2.2$ & $108.07/106$ & $33.3 \pm 2.1$ & $19.0 \pm 2.2$  \\
GH 142c & $0.10\pm 0.01$ & $6.02 \pm 1.04$ & $1.45 \pm 0.11$ & $27.2 \pm 2.4$ & $177.26/136$ & $26.3 \pm 2.3$ & $16.4 \pm 2.4$  \\
GH 142d & $0.085 \pm 0.005$ & $6.9 \pm 0.7$ & $1.58 \pm 0.03$ & $36.2 \pm 0.8$ & $278.90/231$ & $29.7 \pm 0.55$ & $17.7 \pm 0.6$  \\
GH 181a & $0.12 \pm 0.02$ & $0.53 \pm 0.13$ & $1.84 \pm 0.07$ & $4.4 \pm 0.3$ & $104.45/91$ & $2.79 \pm 0.08$ & $1.44 \pm 0.07$  \\
GH 181b & $0.12 \pm 0.02$ & $0.64 \pm 0.19$ & $1.84 \pm 0.11$ & $4.3 \pm 0.4$ & $54.68/48$ & $2.81 \pm 0.14$ & $1.41 \pm 0.14$  \\
GH 211b & $0.10 \pm 0.03$ & $0.15_{-0.06}^{+0.17}$ & $1.28 \pm 0.36$ & $0.37 \pm 0.11$ & $4.57/5$ & $0.46 \pm 0.09$ & $0.30 \pm 0.07$ \\
GH 211d &$0.09 \pm 0.02$ &$0.52 \pm 0.16$ &$1.82 \pm 0.1$ & $3.57 \pm 0.27$ & $56.04/59$ & $2.37 \pm 0.15$ & $1.21 \pm 0.16$ \\
GH 213 &$0.11 \pm 0.03$ & $0.36 \pm 0.16$ & $1.97 \pm 0.17$ & $2.3 \pm 0.3$ & $15.43/28$ & $1.29 \pm 0.08$ & $0.60 \pm 0.09$ \\
\hline
\end{tabular}
\end{center}
\small
Note.--- $kT$ is given in keV.  Flux is given in the units of $10^{-13}$ erg  cm$^{-2}$ s$^{-1}$.  $N_{\rm H}$ is given in Table~\ref{tab:sample}. The normalization of the blackbody component is given as $L_{39}/D_{10}^{2}$, where $L_{39}$ is the luminosity of the source in units of 10$^{39}$ erg s$^{-1}$ and $D_{10}^{2}$ is the distance to the source in units of 10 kpc. The normalization of the power-law is the same as is defined in Table 2.
\end{minipage}
\end{table*}

\subsection{Type II objects}\label{sec:type2}

As mentioned above, fits to GH 47 and GH 94 show different spectral shapes compared to the other objects.  In the case of GH 47, both a simple power-law and a power-law plus blackbody model produce extremely poor fits (reduced-$\chi^2 > 4.1$ in both cases).  For GH 94 (also known as NGC 3259), both spectra are of low quality, with the PN spectra having 250 counts in GH94a and 200 counts in GH94b (in the $0.3 - 10$ keV range).  Even so, fits with a simple power-law are poor (reduced-$\chi^2 > 3.9$ in both cases).  An acceptable fit can be achieved if fitting a power-law plus blackbody model.  However, in this case the power-law index for both spectra is $\Gamma < 0$, far from the standard index for type-I objects.  Visual inspection of the spectra show that they resemble type II AGN X-ray spectra rather than type I spectra.  Note that GH 94a was previously analyzed by \citet{thornton09}, who also found the same problems fitting this spectrum with simple models (our results are consistent with theirs).  Their best-fit was a model including partial covering absorption, similar to the type of model we apply to GH 47 below.

\begin{figure}
\centering
\includegraphics[angle=270,width=8.4cm]{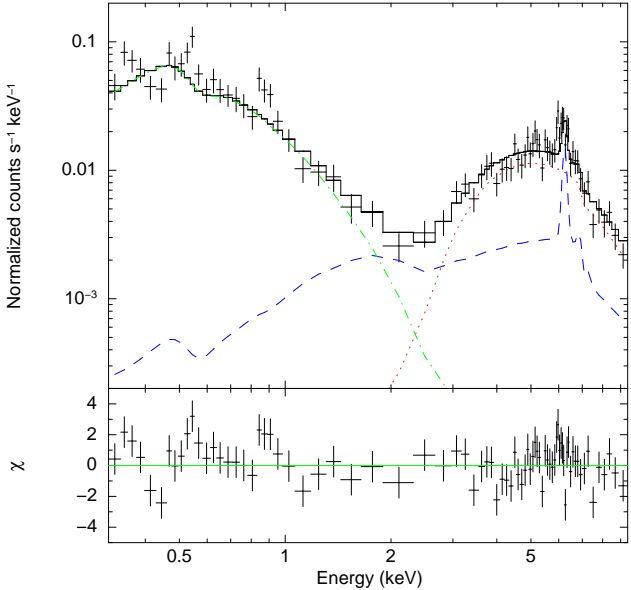}
\caption{The PN spectrum of GH~47.  The solid (black) line indicates the best-fitting type-II AGN model (see text for details).  The blue dashed line indicates the distant reflection (pexmon) component, the red dotted line is the intrinsic power-law component and the green dash-dotted line is the scattered power-law component.  The bottom panel shows the normalized residuals.}
\label{fig:type2}
\end{figure}

The spectra of GH 94 are too poor to require more complex modeling, however, for GH 47 we fitted a model appropriate for type II AGN, consisting of {\sc phabs(power-law + pexmon + zphabs(power-law))}.  In this model the first absorber is due to Galactic absorption, the first power-law is due to scattered emission, the pexmon component accounts for reflection from the torus and the absorbed power-law accounts for the near Compton-thick absorption of the intrinsic power-law emission from close to the black hole.  This model provides a significantly improved fit, indicating GH47 is likely a type-II AGN.  We give the best-fitting spectral parameters in Table~\ref{tab:type2}, and show the best-fitting model along with the PN spectrum in Figure~\ref{fig:type2}.

\begin{table*}
\caption{Best fitting spectral parameters for GH 47}
\label{tab:type2} 
\begin{center}
\begin{tabular}{llc}
\hline
Component & Parameter & Value \\
\hline
phabs & $N_{\rm H}$ ($10^{22}$~cm$^{-2}$) & $0.17\pm0.06$  \\
pexmon & $\Gamma$ & $2.0\pm0.2$ \\
 & $E_{\rm fold}$ (keV) & 300 (fixed) \\
 & $i$ ($^\circ$) & 60 (fixed) \\
 & Normalization & $\left(1.2_{-0.5}^{+0.8}\right) \times10^{-3}$ \\
power-law (scattered) & $\Gamma$ & $4.1_{-0.5}^{+0.3}$ \\
 & Normalization & $(2.8_{-0.4}^{+0.7})\times10^{-5}$   \\
zphabs &  $N_{\rm H}$ ($10^{22}$~cm$^{-2}$) & $21\pm3$ \\
power-law (intrinsic) & $\Gamma$ & = pexmon value\\
 & Normalization & $(7.9_{-1.7}^{+4.5})\times10^{-4}$ \\
 0.5 -- 10 keV flux & ($10^{-13}$ erg cm$^{-2}$ s$^{-1}$) & $9.9^{+1.3}_{-6.3}$ \\ 
\hline 
 $\chi_\nu^2$ (dof) & & 1.48 (128) \\
\hline
\end{tabular}
\end{center}
Note --- the reflection scaling factor is set to $-1$ to indicate there is no direct component.  Abundances were all set to solar values, and $z$ was fixed to 0.0254 in both {\sc pexmon} and {\sc zphabs}. The normalization of the pexmon component is the photon flux at 1 keV. The normalization of the power-law components is defined in Table 2.
\end{table*}

\subsection{Variability: calculating normalized excess variance}
Another objective of studying this sample of low-mass AGN was to characterize their variability and compare them with more massive counterparts.  There are well known anti-correlations between measures of variability and black hole mass \citep[see, e.g.][and references therein]{ponti12,kelly13,mchardy13}.  \citet{miniutti09} found evidence that the four GH objects they looked at extended the relationship to lower masses.  Here, we added 10 GH objects to the low-mass end as well as additional data for one of the objects in \citet{miniutti09}.

\begin{figure}
\centering
\includegraphics[angle=270,width=8.4cm]{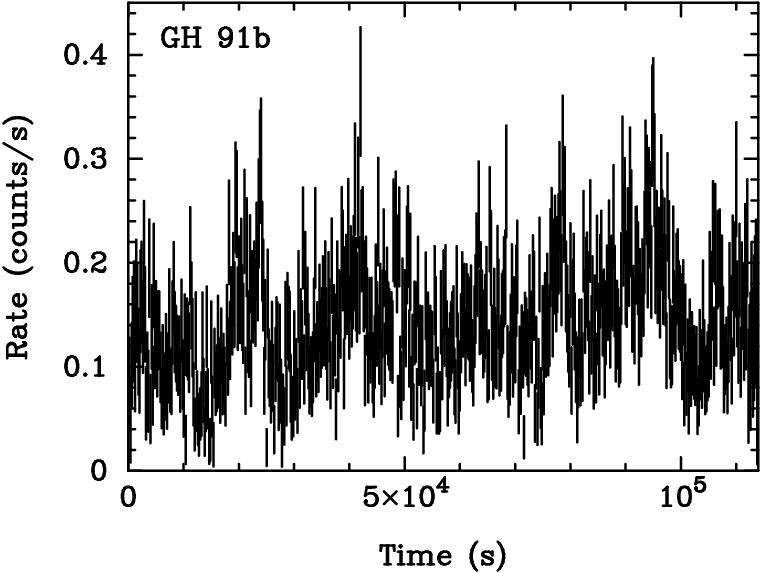}\\[4ex]
\includegraphics[angle=270,width=8.4cm]{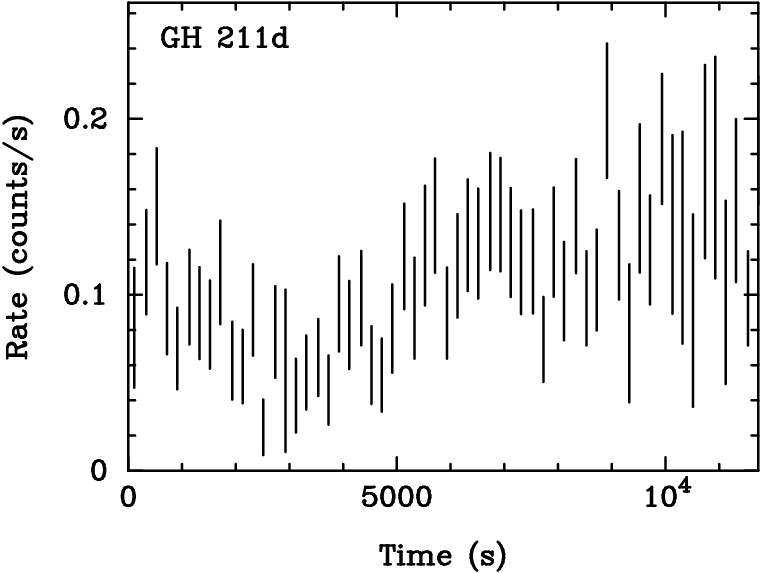}
\caption{0.2 -- 10 keV background-subtracted lightcurves from 2 objects in our sample, GH 91b (top) and GH 211d (bottom).}
\label{fig:lc}
\end{figure}

We used PN lightcurves in the $0.2 - 10$ keV range, binned to 200s.  The lightcurves exhibit rapid, short timescale variability (see the two example lightcurves in Figure~\ref{fig:lc}).  We proceeded to do a comparison of SMBH and IMBH variability by calculating the normalized excess variance, $\sigma_{\rm NXS}^2$.  The process of calculating the excess variance can be found in \citet{vaughan03}.  The normalized excess variance is just the excess variance divided by the square of the average value of the lightcurve ($\sigma_{\rm NXS}^2 = \sigma_{\rm XS}^2 / \bar{x}^2$). The excess variance is obtained from subtracting expected measurement errors from the standard variance.  For the purposes of variability studies we omitted GH 47 and GH 94 from the sample given we find evidence that these objects are Type II AGNs (though see comments on the variability of GH 47 at the end of Section~\ref{sec:var_mbh}). For GH 181a, we used lightcurves that contained the high background periods in order to calculate the excess variance.  Filtering for the background would cause multiple gaps of missing data that would have given a misrepresentation of the objects variance.  

We chose 10 ks time intervals to calculate the excess variance \citep[note that][used 20 ks time intervals]{miniutti09} so that we could use data from each of the objects in our sample (the observations were generally shorter than 20 ks).  We also recalculated the excess variance for the GH objects in \citet{miniutti09} on 10 ks time intervals in order to directly  compare with the newly analyzed GH objects presented here.  For those that had observations longer than 10 ks, their data was broken up into 10 ks time intervals and the excess variance was averaged for that object.   Segments where the signal-to-noise ratio was low and the measured excess variance was less than zero (i.e. when the measurement errors are larger than the standard variance) were not included in the averaging. Table~\ref{tab:exvar} gives the normalized excess variances for all GH objects analyzed here.  GH 79 had a highly unconstrained value, and thus we do use it here.

The statistical uncertainty in the normalized excess variance was calculated using the formulae in \citet{vaughan03}, who investigate the error in the measured excess variance through Monte Carlo simulations.  Importantly, from the Monte Carlo simulations they find that the normalized excess variance is an unbiased estimator of the intrinsic variance, even when the signal-to-noise ratio is relatively low.  In addition to the statistical uncertainty there is also a stochastic uncertainty as the variance of the lightcurve depends upon when it is observed.  \citet{vaughan03} also examine this in detail, and we follow the \citet{ponti12} method for including this in the total uncertainty in the normalized excess variance (see their Appendix A, though we calculate the 1$\sigma$ error rather than the 90\% confidence level).  Note that in almost all cases the stochastic uncertainty completely dominates over the statistical uncertainty.

\begin{table*}
\caption{BH mass and excess variance for the GH objects}
\label{tab:exvar} 
\begin{minipage}{167mm}
\begin{center}
\begin{tabular}{lccccccc}
\hline
Object & $\log (M_{\rm BH}/ M_{\odot})$ & $\log (\sigma_{\rm NXS}^2)$ & $\log (L_{\rm bol, X})$ & $L_{\rm bol,X}/L_{\rm Edd}$ & $\log (L_{\rm bol,  H\alpha})$ & $L_{\rm bol, H\alpha}/L_{\rm Edd}$ & $L_{\rm bol,X}/L_{\rm bol, H\alpha}$ \\
\hline
GH 11 & 6.26 & $-1.36\pm0.51$ & 43.71  & 0.22 & 43.89 & 0.34 & 0.65  \\
GH 18 & 5.52 & $-1.55\pm0.41$ & 43.29 & 0.43 & 42.98 & 0.23 & 2.09 \\
GH 49 & 6.03 & $-1.40\pm0.30$ & 43.57 & 0.28 & 43.61 & 0.30 & 0.93 \\
GH 78 & 6.26 & $-1.72\pm0.14$ & 43.71 & 0.22 & 43.90 & 0.34 & 0.65 \\
GH 91 & 5.78 & $-1.28\pm0.55$ & 43.12 & 0.18 & 43.47 & 0.39 & 0.45 \\
GH 112 & 5.88 & $-1.34\pm1.05$ & 44.11 & 1.36 & 43.58 & 0.40 & 3.44 \\
GH 116 & 5.39 & $-0.82\pm0.88$ & 42.18 & 0.049 & 42.56 & 0.12 & 0.41\\
GH 126 & 6.20 & $-1.80\pm0.45$ & 43.78 & 0.30 & 44.00 & 0.50 & 0.60 \\
GH 138 & 6.03 & $-1.36\pm0.46$ & 43.52 & 0.24 & 43.56 & 0.27 & 0.90 \\
GH 142 & 5.78 & $-1.61\pm1.01$ & 42.93 & 0.11 & 42.91 & 0.11 & 1.04 \\
GH 171 & 6.29 & $-1.82\pm0.49$ & 44.11 & 0.53 & 44.03 & 0.43 & 1.09 \\
GH 181 & 5.86 & $-0.79\pm0.75$ & 42.36 & 0.025 & 42.94 & 0.096 & 0.26 \\
GH 211 & 5.82 & $-0.91\pm1.09$ & 42.15 & 0.017 & 43.30 & 0.24 & 0.071\\
GH 213 & 5.90 & $-1.41\pm1.63$ & 42.37 & 0.023 & 43.37 & 0.24 & 0.10\\
\hline
\end{tabular}
\end{center}
Note.--- $L_{\rm bol, X}$ is the bolometric luminosity given in erg s$^{-1}$ based on the $2-10$ keV flux, applying the bolometric correction of \citet{marconi04}.  $L_{\rm bol, H\alpha}$ is the bolometric luminosity given in erg s$^{-1}$ based on the H$\alpha$ luminosity from \citet{Greene07} and is given here for comparison.  GH names refer to \citet{Greene07}.  Note that GH 11, GH 126, GH 171, and GH 181 listed here are sometimes alternatively known as GH 1, GH 8, GH12, and GH14 when using the \citet{greene04} rather than \citet{Greene07} ID numbers.
\end{minipage}
\end{table*}

\subsection{Variability and black hole mass}\label{sec:var_mbh}

The most comprehensive study to date of black hole mass versus normalized excess variance was performed by \citet{ponti12}, where they calculated the normalized excess variance for a large sample of variable AGN observed by {\it XMM-Newton}.  \citet{ponti12} calculate the normalized excess variance in the $2-10$ keV band.  However, in our sample of GH objects many of the sources are background dominated above 2 keV.  In order to maximize the number of sources we could obtain a measurement of the normalized excess variance we therefore used lightcurves in the $0.2 - 10$ keV range.  We cannot directly compare our results with those of \citet{ponti12}.  We therefore calculated the normalized excess variance for the sample of Seyfert 1s used in \citet{miniutti09} using $0.2 - 10$ keV lightcurves and 200s bins (see Table~\ref{tab:S1sample}).  We note that from the \citet{miniutti09} sample we removed two sources (IC~4329A and PG~1211+143) as they are determined to have low-quality black hole mass estimates by \citet{peterson04}.  A third source, Fairall 9, had a very large stochastic uncertainty in the normalized excess variance, and so we do not use that object here either. 

In order to make the most robust comparison between normalized excess variance and black hole mass we updated the GH black hole masses using the latest scaling relation involving the H$\alpha$ luminosity and FWHM from \citet{reines13} (their equation 5), using $\epsilon = 1.0775$.  \citet{reines13} use the most recent $R$-$L$ scaling relation from \citet{bentz13} following the approach of \citet{greene04,Greene07}. $\epsilon = 1.0775$ corresponds to the mean virial factor $\langle f \rangle = 4.31\pm1.05$ determined by \citet{grier13} from recalibrating reverberation masses to the $M-\sigma$ relation, and using $\epsilon = f/4$ from assuming that $\sigma = V_{\rm FWHM} / 2$ \citep{onken04}. Table~\ref{tab:exvar} gives the updated GH black hole masses.  Note that the updated masses are systemically slightly larger than the original GH masses, but it is not a large effect.

We also use the most up-to-date masses for the Seyfert 1 sample.  For the objects whose masses have been measured directly from reverberation mapping, we use the most recent masses from \citet{grier13}, where available.  For those reverberation mapped AGN not in \citet{grier13}, we use the \citet{peterson04} virial product, to calculate the black hole mass using the \citet{grier13} $f$ value, except for NGC 4395 where we use the \citet{peterson05} virial product.  The other AGNs have their masses estimated via reverberation-based scaling relations, using the 5100\AA\ luminosity to estimate the broad line region radius (from the $R$-$L$ relation) and combining the radius with the FWHM of the H$\beta$ broad emission line in order to get the mass.  We obtained the $\lambda L_\lambda (5100$\AA) and FWHM(H$\beta$) values for each object from the literature, and used the most recent $R$-$L$ relation of \citet{bentz13} in order to estimate $R_{\rm BLR}$ \citep[their `clean' fit, also used in determining the H$\alpha$ relations of][]{reines13}.  Of all the sources, for only two could we not find literature values for $\lambda L_\lambda (5100\AA)$ and FWHM(H$\beta$).  For those sources (1H~0707$-$495 and IRAS~13224$-$3809) we use literature values for black hole mass estimates.  All masses for the Seyfert 1 sample are given in Table~\ref{tab:S1sample}.

\begin{table*}
\caption{Mass and variability for the Seyfert 1 sample}
\label{tab:S1sample}
\begin{tabular}{lcccl}
\hline
Name & $\log (M_{\rm BH}/ M_{\odot})$ & $f_{\rm Edd}$ & $\log\sigma_{\rm NXS}^2$ & Mass reference \\
\hline
1H 0707$-$495 &   6.37 &      0.17 &	 $-1.14\pm0.30$ & \citet{zhouwang05} \\
Ark 120 &   7.98 &      0.25 & $-4.05\pm1.36$ & \citet{grier13}\\
Ark 564 &   6.65 &      1.07 & $-1.70\pm 0.38$ & \citet{botte04}\\    
HE~1029$-$1401 &   9.31 &   0.067 & $-3.30\pm0.69$ & \citet{mclure01} \\
IRAS~13224$-$3809 &   6.76  &    0.16 &	 $-0.94\pm0.37$ & \citet{zhouwang05} \\   
I Zw 1 &   7.45 &      0.76 &  	$-2.20\pm0.53$ & \citet{botte04} \\   
MCG$-$6$-$30$-$15 &  6.87 &  0.090 &	$-1.58\pm0.45$ & \citet{mchardy05} \\
Mrk 110 &   7.32 &    0.99 &  $-3.75\pm0.86$ & \citet{grier13} \\
Mrk 335 &   7.03 &    0.26 & $-2.00\pm0.69$ & \citet{peterson04} \\
Mrk 478 &   7.55 &    0.34 & $-2.04\pm0.43$ & \citet{grupe04} \\ 
Mrk 509 &   7.95 &    0.39 & $-3.98\pm0.53$ &  \citet{grier13}\\     
Mrk 766 &   6.10 &    0.91 & $-1.75\pm0.33$ &  \citet{grier13}\\
Mrk 841 &   8.24 &    0.053 & $-2.99\pm0.90$ & \citet{grupe04} \\
MS 2254$-$36 &   6.89 &  0.22 & $-1.99\pm0.49$ & \citet{grupe04}\\
NGC 3783 &   7.26 &  0.090 &	$-2.63\pm0.44$ &  \citet{grier13} \\
NGC 4051 &   6.32 &  $6.9\times10^{-3}$ & $-0.99\pm0.35$  &  \citet{grier13}\\
NGC 4151 &   7.65 &  0.013 &	$-3.34\pm0.55$ &  \citet{grier13} \\
NGC 4395 &   5.44 &  $3.5\times10^{-3}$ & $-0.74\pm0.43$  & \citet{peterson05} \\
NGC 4593 &   6.94 &   0.11 &  $-2.43\pm0.34$  &  \citet{grier13}\\
NGC 5548 &   7.76 &   0.088 & $-3.54\pm0.76$  &  \citet{grier13} \\
NGC 7469 &   7.30 &   0.16 &  $-3.08\pm0.33$  &  \citet{grier13}\\
Ton~S180 &   7.00 &    0.94 & $-1.93\pm0.33$  & \citet{grupe04} \\
WAS 61 &   6.96 & 0.55 &	$-2.38\pm0.39$ & \citet{grupe04} \\ \hline
\end{tabular}
\end{table*}

We compare the GH objects to the Seyfert 1 sample in order to see whether the anti-correlation between black hole mass and normalized excess variance holds at low-mass, as suggested from the 4 GH objects studied by \citet{miniutti09}.    Fitting the Seyfert 1 sample with a simple linear relation \citep[using a linear bisector as in][]{ponti12}, we  find 
\small
\begin{equation}
\log(\sigma_{\rm NXS}^2) = (-1.23\pm0.22)\log\left(\frac{M_{\rm BH}}{10^7~\rm M_\odot}\right) - (2.15\pm0.12)
\end{equation}
\normalsize
for the sample we are using.  Note that this is consistent with the similar fits in \citet{ponti12} where the excess variance in calculated in the 2--10 keV band.

\begin{figure}
\centering
\includegraphics[angle=270,width=8cm]{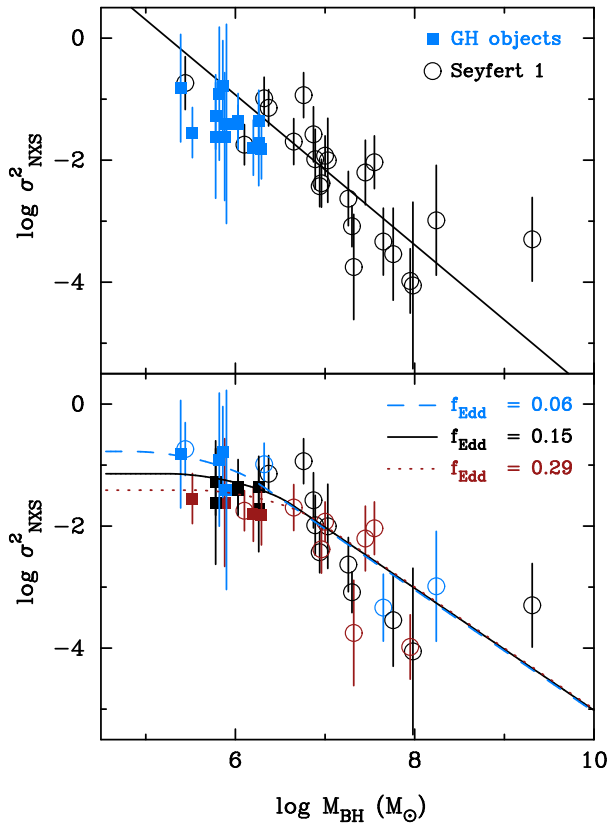}
\caption{Normalized excess variance ($\sigma^2_{\rm NXS}$) versus black hole mass.  {\it Top:} The black open circles indicate the sample of Seyfert 1 objects, while the blue filled squares indicate the GH objects.  The line is the best linear-bisector fit to the Seyfert 1 sample. {\it Bottom:} Colors of the data points indicate the range of $f_{\rm Edd}$, with blue indicating $f_{\rm Edd} < 0.06$, black $0.06 \leq f_{\rm Edd} \leq 0.29$ and red $f_{\rm Edd} > 0.29$.  The lines are plotted for $f_{\rm Edd}$ = 0.06 (dashed blue), $f_{\rm Edd} = 0.15$ (solid black) and $f_{\rm Edd} = 0.29$ (dotted red).  These are the 33rd, 50th and 67th percentile of the $f_{\rm Edd}$ distribution of the sample.}
\label{fig:exvar}
\end{figure}

In the top panel of Figure~\ref{fig:exvar} we compare the normalized excess variances of the GH black holes (blue squares) to those from the Seyfert 1 sample (black circles).  Broadly speaking, it appears that the GH objects show that this relation extends to low masses.  However, we note that all 14 of the GH objects lie below the best-fit line quoted above.  If measurements are equally likely to lie above and below the line, this would happen $0.012$\% of the time (using a two-sided binomial test), and hence we can rule out an extrapolation of the linear relation to these data at  $3.8\sigma$ confidence.

Such an offset for low-mass objects is expected if all AGN have the same underlying power spectrum with a slope of $-2$ at high frequencies that breaks to a slope of $-1$ at low frequencies, with a break frequency, $\nu_{\rm br}$, that scales with mass \citep{papadakis04,oneill05,gonzalez11,ponti12,kelly13}.  In calculating the excess variance we only probe timescales between the binning time of the lightcurves (200s here) and the length of the lightcurve segments (10 ksec here).  For black holes with masses significantly above $10^6$~M$_\odot$, the break frequency is low enough that it occurs on timescales longer than the 10 ksec lightcurve length.  The lightcurves are therefore just sampling variability occuring on the part of the power spectrum with a slope of $-2$.  However, once masses get below a few times $10^6$~M$_\odot$, $\nu_{\rm br}$ corresponds to a timescale shorter than the length of the lightcurve, and we start to also probe the power spectrum with a slope of $-1$.  This changes the expected black-hole-mass--excess-variance relation, which should flatten from a slope of $-1$ at higher masses to a slope of 0 below $\sim10^6$~M$_\odot$ \citep{papadakis04,oneill05,gonzalez11,ponti12}.  Having all 14 GH objects lie below a simple linear relation determined from more massive objects is therefore consistent with expectations of AGN having a universal power spectrum that scales with black hole mass.

We proceed to fit the black-hole-mass--excess-variance relation with models that assume a universal power spectrum.  We follow the prescription of \citet{gonzalez11} as modified by \citet{ponti12} to allow for further dependencies on mass accretion rate.  We describe the model below.  Note that in previous work, authors tend to refer to the mass accretion rate as a fraction of the Eddington ratio, $\dot{m}_{\rm Edd}$, which is usually determined from $L_{\rm bol}/L_{\rm Edd}$.  We prefer to refer to this as the Eddington fraction, $f_{\rm Edd}$, since, there is a correction for efficiency to go from $L_{\rm bol}$ to $\dot{m}$ that is not included when using $\dot{m}_{\rm Edd} = L_{\rm bol}/L_{\rm Edd}$.

First, we assume that the break frequency scales with both $M_{\rm BH}$ and $f_{\rm Edd}$ \citep{mchardy04,uttley05,mchardy06,kording07}.  As shown by \citet{mchardy06}:
\begin{equation}
\nu_{\rm br} = 0.003 f_{\rm Edd} (M_{\rm BH}/10^6 \; {\rm M}_\odot)^{-1} \; \rm Hz
\end{equation}
Following \citet{gonzalez11} we then calculate $\sigma^2_{\rm NXS}$ using:

\footnotesize
\begin{equation}
\sigma^2_{\rm NXS} = \left\{
\begin{array}{ll}
  C \nu_{\rm br}(\nu^{-1}_{\rm min} - \nu_{\rm max}^{-1}), & (\rm{if} \; \nu_{\rm br} < \nu_{\rm min}) \\
  C\left[\ln\left(\frac{\nu_{\rm br}}{\nu_{min}}\right) - \frac{\nu_{\rm br}}{\nu_{\rm max}} + 1 \right], & (\rm{if} \; \nu_{\rm min} < \nu_{\rm br} < \nu_{\rm max}) \\
  C\ln\left(\frac{\nu_{\rm max}}{\nu_{\rm min}}\right), & (\rm{if} \; \nu_{\rm br} > \nu_{\rm max})
\end{array}
\right.
\end{equation}
\normalsize
\citet{ponti12} argue that the presence of objects with similar mass but quite different values of $\sigma^2_{\rm NXS}$ suggest a dependence of the normalization of the power spectrum on $f_{\rm Edd}$.  We therefore parameterize our model similarly: 
\begin{equation}
C = A f_{\rm Edd}^{-\beta}
\end{equation}
which is referred to as Model B by \citet{ponti12}.  In fitting this to the sample, then, $A$ and $\beta$ are the free parameters, while $M_{\rm BH}$, $\sigma^2_{\rm NXS}$ and $f_{\rm Edd}$ are the observables.

To fit this model we need to determine the bolometric luminosity of the objects in our sample in order to estimate $f_{\rm Edd}$.  \citet{ponti12} use the bolometric correction of \citet{marconi04} in order to go from $2 - 10$ keV X-ray luminosity to bolometric luminosity.  For consistency, we therefore also take this same approach to determine the bolometric luminosities of the GH objects and Seyfert 1 sample.  In Table~\ref{tab:exvar} we give the bolometric luminosity based on the average $2 -10$ keV luminosity of each GH object in our sample, along with the Eddington fraction.   We calculate X-ray luminosities assuming $H_0 = 70$ km s$^{-1}$ Mpc$^{-1}$, $\Omega_m = 0.3$, $\Omega_\Lambda = 0.7$.  For comparison, we also give the bolometric luminosity based on the H$\alpha$ luminosity \citep[taken from][]{Greene07}, the associated Eddington fraction, and the ratio of the Eddington fraction based on the X-ray luminosity to the Eddington fraction based on the H$\alpha$ luminosity.  Eddington fractions for our Seyfert 1 sample are given in Table~\ref{tab:S1sample} and are also obtained from using the bolometric correction of \citet{marconi04} on the $2-10$ keV unabsorbed fluxes.  We obtain the $2-10$ keV unabsorbed fluxes through spectral fitting of all observations in our sample.

We fitted the above model for the mass--variability relation to the full sample of objects using the X-ray determined values of $f_{\rm Edd}$.  We assume that $\sigma^2_{\rm NXS}$ is the dependent variable in the fits.    We determine the uncertainties in $A$ and $\beta$ from 10,000 Monte Carlo simulations, and take the median of the distribution as the best-fitting values and the 68\% interval as our uncertainties. We get a best-fit with $A = (3.5\pm1.0)\times10^{-3}$, $\beta = 0.94 \pm 0.13$.

To graphically demonstrate the best-fit (we cannot simply plot a best-fit line since each object has a different $f_{\rm Edd}$), we split the sample into 3 sub-samples based on their $f_{\rm Edd}$.  In the bottom panel of Figure~\ref{fig:exvar}, the objects in the bottom third of the distribution ($f_{\rm Edd} < 0.06$) are shown in blue, those in the middle third are shown in black ($0.06 \leq f_{\rm Edd} \leq 0.29$) and those in the upper third ($f_{\rm Edd} > 0.29$) are shown in red.  The three lines represent the boundaries between these regions and the median value.  It can clearly be seen that above approximately $10^6$~M$_\odot$ all the models show approximately the same relation.  Below this mass the models all flatten and there is a small dispersion between the lowest and highest $f_{\rm Edd}$ models, as expected by the model.

\begin{figure}
\includegraphics[angle=270,width=8.4cm]{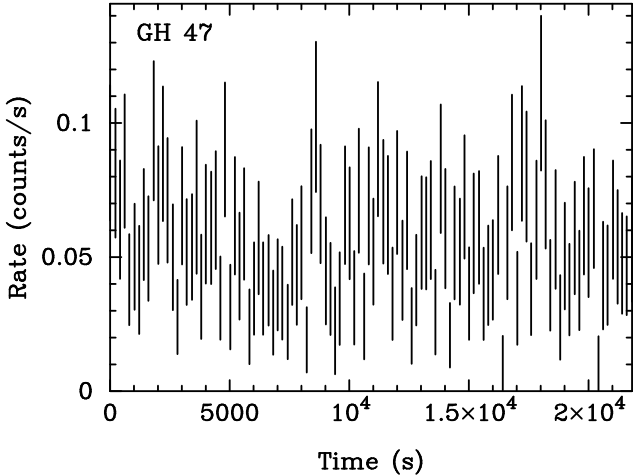}
\caption{0.2 -- 2 keV background-subtracted lightcurve of the type-II object GH 47.}
\label{fig:lc_gh47}
\end{figure}

Finally, we consider the variability in the type-II object, GH 47.  We would expect the variability of the soft emission (dominated by the scattered power-law component) to have little variability, and that is what we find.  In Figure~\ref{fig:lc_gh47} we show the 0.2 -- 2 keV lightcurve of GH 47.  The normalized excess variance across the 0.2 -- 10 keV band (in order to compare to the other GH objects) is $\log \sigma^2_{\rm NXS} = - 2.9$, significantly smaller than all the other GH objects, as would be expected if it is a type II object.

\subsection{Long timescale variability}

In addition to short time-scale variability (shorter than the length of observations), there is also variability exhibited by our sample over a long period of time (between observations: months -- years).  The fluxes originally detected by the \emph{ROSAT} All Sky Survey (RASS) are listed in Table~\ref{tab:rosat}.  The fluxes from the time that they were observed by \emph{XMM-Newton} are listed in the following column.  The difference in the observed fluxes over these extended time periods between observation shows long-term variability, not just between {\it ROSAT} and {\it XMM-Newton} observations, but also between individual {\it XMM-Newton} pointings.  Several objects' fluxes have changed by a factor of four between the time that \emph{ROSAT} and \emph{XMM-Newton} had observed them.  Between the four observations of GH 211 alone the flux changed by a factor of almost 10. 

\begin{table}
\caption{Flux comparison in the 0.5-2.0 keV range}
\label{tab:rosat}
\begin{center}
\begin{tabular}{lccc}
\hline
\multicolumn{1}{c}{Object}
&\multicolumn{1}{c}{ROSAT}
&\multicolumn{1}{c}{Date of XMM obs}
&\multicolumn{1}{c}{XMM-Newton}\\
\hline
GH 18
&$2.2 \pm 0.8$
&17/01/2012
&$1.80 \pm 0.04$\\
GH 78
&$10.7 \pm 3.7$
&30/10/2005
&$4.94 \pm 0.06$\\
GH 112
&$3.6 \pm 2.1$
&03/11/2011
&$2.11 \pm 0.29$\\
GH 138a
&$2.3 \pm 1.9$
&29/05/2011
&$0.55 \pm 0.08$\\
GH 138b
&
&05/11/2011
&$2.03 \pm 0.04$\\
GH 142a
&$2.3 \pm 1.5$
&19/06/2007
&$10.4 \pm 0.1$\\
GH 142c
&
&03/05/2011
&$10.8 \pm 0.3$\\
GH 142d
&
&20/06/2011
&$7.56 \pm 0.22$\\
GH 142e
&
&30/10/2011
&$9.13 \pm 0.09$\\
GH 181a
&$2.4 \pm 2.0$
&18/08/2005
&$1.07 \pm 0.02$\\
GH 181b
&
&16/08/2011
&$1.09 \pm 0.03$\\
GH 211a
&$5.2 \pm 3.3$
&17/08/2007
&$0.092 \pm 0.013$\\
GH 211b
&
&19/08/2007
&$0.12 \pm 0.01$\\
GH 211c
&
&16/08/2011
&$0.32 \pm 0.02$\\
GH 211d
&
&17/01/2012
&$0.87 \pm 0.02$\\
GH 213
&$2.2 \pm 0.7$
&28/08/2011
&$0.53 \pm 0.02$\\
\hline
\end{tabular}
\end{center}
Note.--- Flux is given in units of $10^{-13}$ erg  cm$^{-2}$ s$^{-1}$.  There is a significant change in the flux between the time \emph{ROSAT} and \emph{XMM-Newton} took their observations. \emph{ROSAT} detection flux values are obtained from Table 4 in \citet{Greene07}. 
\end{table}

\section{Discussion}\label{sec:discuss}

Using observations with {\it XMM-Newton}, we have determined the X-ray spectral and variability properties of a sample of 14 AGNs estimated to have masses $<2\times10^6~M_\odot$ by \citet{Greene07}.  

Of all the sources examined here,  5 had spectra that were not background-dominated at $2-3$ keV.  When fitting the spectra of these 5 objects from $2 - 10$ keV with a simple absorbed power-law and extrapolating back to the $0.3 - 2$ keV range, we found a soft excess was present in all 5 objects.  The power-law indices are close to or within the range consistent with those seen in more massive radio quiet AGNs  \citep[$1.7<\Gamma<2.6$;][]{dewangan08}.  When fitting the soft excess with a blackbody component, the blackbody temperatures were found to be consistent with $0.1-0.2$ keV, as is typically seen in type I AGN \citep{gier04}.  Combining our results with the previous findings of \citet{miniutti09}, we find that a total of 8 GH objects are known to show a soft excess.

The nature of the soft excess is still uncertain, though the fact that the characteristic blackbody temperature is constant over a wide range in black hole mass suggests an atomic origin.  Reflection of hard X-rays off the accretion disk can successfully fit the soft excess \citep{crummy06,walton13}, but may not be sufficient or required in every instance \citep{lohfink12,matt14}.  An alternative suggestion for the origin of the soft excess involves optically thick Comptonized disc emission \citep{done12}, which should be most prominent in lower-mass objects due to their hotter discs.  The soft excesses present in a significant fraction of the low-mass AGN studied here would be a good test of these models if a particularly high S/N spectrum was obtained, especially because the \citet{done12} model should be most prominent in lower-mass objects.  We note that the spectra are all relatively smooth at low energies, with no spectrum statistically requiring additional emission lines.  In the reflection origin for the soft excess, this would imply these objects are rapidly spinning (in order to smooth out reflection features).

Two objects in the sample, GH 47 and GH 94, both show spectra that cannot be simply fit by a power-law plus a blackbody.  Those objects displayed spectra consistent with being classed as type II AGN.  Moreover, the lightcurve of GH 47 has significantly less variability than the other GH objects, as expected from our type II classification.   While these two objects were initially classed as type I AGN optically (based on the presence of a broad H$\alpha$ line), they appear to be type II from X-ray observations.  We note, however, that a follow-up Magellan observation of GH 47 by \citet{xiao11} found that the peak amplitude of the H$\alpha$ line was not as significant as other type I AGN in comparison to the rms deviation of the continuum-subtracted spectrum. These authors therefore flagged this object as having only a possible broad H$\alpha$ line which makes the optical type I classification insecure. Additionally,  \citet{thornton09} classify GH 94 as a type II based on the weak broad H$\alpha$ emission line. Thus, the type I optical classification of both these objects is questionable.  However, if these two objects are type I AGN optically yet type II in the X-ray regime, then this adds to the sample of AGN where optical and X-ray classifications differ \citep[e.g., see the discussion in][]{matt02}.  
 
We also studied the X-ray variability properties of these GH objects.  A comparison of the fluxes of the objects from the ROSAT All Sky Survey and multiple \emph{XMM-Newton} observations shows long-term variability (timescale of months to years).  We also detected significant short-term variability (hundreds to thousands of seconds) during the observations.  It has long been known that short timescale variability amplitude  scales inversely with black hole mass \citep[see][and references therein]{ponti12,kelly13}.  A similar inverse scaling holds even when fitting for the overall normalization of the X-ray power spectrum \citep{mchardy13}.  This is expected based simply on the fact that the size scale for the system scales linearly with black hole mass (the gravitational radius, $r_g = GM/c^2$).  If this is the dominant effect, then more massive black holes have bigger accretion disks and thus will be less variable on shorter timescales.  Such inverse scaling with black hole mass has been observed when using either the normalized excess variance \citep[e.g.][]{ponti12} or the rate of stochastic variability power \citep{kelly13} as a measure of the variability amplitude.  The scatter in the relation is also quite small ($\sim0.3$ dex), indicating that measures of variability are a promising way to estimate black hole masses.

No study of variability amplitude versus mass has included a significant number of black holes at the low-mass end, thus it was not clear to what extent this relation can be extrapolated to lower masses, although, \citet{miniutti09}, who studied 4 GH objects, found that they appeared to extend the relation to lower masses.  Here, we compare a total of 14 GH objects with a sample of 23 Seyfert 1s.  We find that the GH objects do have high normalized excess variance values, as expected for lower mass AGN, but that all 14 objects lie below the best-fitting log-linear relation fit to more massive Seyfert 1s, ruling out a simple extrapolation of that relation at $3.8\sigma$ confidence.  This is expected in the case of a universal power spectrum with a break frequency that scales with mass \citep{papadakis04,oneill05,gonzalez11,ponti12,kelly13}.  For low mass objects the break frequency becomes larger than the frequencies probed by the lightcurve.  We proceed to fit the black hole mass versus excess variance relation with models taking this into account, and find that the data support that these are low mass objects displaying variability that follows a universal power spectrum, supporting previous work showing the scaling of the power spectrum between black hole X-ray binaries and AGN \citep{mchardy06}.

As discussed by previous authors \citep[e.g.][]{ponti12,kelly13},  because the black hole mass -- excess variance relation flattens off for masses below approximately $10^6$~M$_\odot$ it cannot be used to directly measure the mass of objects below this value.  However, the fact that we see the relation flatten demonstrates that the AGN have masses below approximately this value.  In order to use variability to independently estimate the black hole masses of these objects it would require using a method such as the rate of stochastic variability power developed by \citet{kelly13}.  However, it is unclear as to whether the quality of data from the {\it XMM-Newton} observations would be sufficient enough for this kind of analysis.

Not included in the model of how excess variance changes with black hole mass is the effect of black hole spin.  The change in the location of the innermost stable circular orbit between a non-rotating and maximally rotating black hole should have an effect on the variability properties and will add scatter to the black hole mass -- excess variance relation.

In summary,  the spectral and variability properties of the low-mass AGN are similar to more massive Seyferts, with the definite existence of a soft excess in some low-mass AGN, and variability amplitude that is consistent with there being a universal power spectrum for all AGN.

\section*{Acknowledgments}
RL thanks the National Science Foundation for support through a Research Experience for Undergraduates program at Wayne State University (NSF Grant No. PHY-1156651).  This research has made use of data obtained through the High Energy Astrophysics Science Archive Research Center Online Service, provided by the NASA/Goddard Space Flight Center.

\footnotesize{
\bibliographystyle{mn2e}
\bibliography{references}
}

\end{document}